\documentclass[a4paper]{jpconf}
\usepackage{graphicx}
\usepackage{cite}
\usepackage{amsmath}
\usepackage{placeins}
\usepackage{multirow}
\usepackage{tabularx}
\usepackage{color}
\usepackage{hyperref}
\usepackage[table, dvipsnames]{xcolor}
\usepackage{ulem}

\usepackage[utf8]{inputenc}
\usepackage[varg]{txfonts}   % Web of Conferences font
\usepackage{booktabs}

\hypersetup{colorlinks, linkcolor = [rgb]{0,0.0,0.75}, citecolor = [rgb]{0,0.0,0.75}, urlcolor = [rgb]{0,0.0,0.75}}
\definecolor{linkBlue}{rgb}{0,0.0,0.75}

\newcommand{\op}{\mathcal{O}}
\newcommand{\C}{\mathcal{C}}
\newcommand{\xb}{\textbf{\textit{x}}}
\newcolumntype{C}[1]{>{\centering\arraybackslash}m{#1}}

\begin{document}
%%%%%%%%%%%%%%%%%%%%%%%%%%%%%%%%%%%%%%%%%%%%%%%%%%%%%%%%%%%%%%%%%%%%%%%%%%%%%
%
%----------------------------------------------------------------------------
\title{%
Investigation of Doubly Heavy Tetraquark Systems\\ using Lattice QCD
}
%----------------------------------------------------------------------------
\author{%
Martin Pflaumer$^{\ast1}$,
Luka Leskovec$^{2,3}$, 
Stefan Meinel$^4$,
Marc Wagner$^{1,5}$
}
%----------------------------------------------------------------------------
\address{%
$^1$ Goethe-Universit\"at Frankfurt am Main, Institut f\"ur Theoretische Physik, Max-von-Laue-Stra{\ss}e 1, D-60438 Frankfurt am Main, Germany \\
$^2$ Thomas Jefferson National Accelerator Facility, Newport News, VA 23606, USA \\
$^3$ Department of Physics, Old Dominion University, Norfolk, VA 23529, USA \\
$^4$ Department of Physics, University of Arizona, Tucson, AZ 85721, USA \\
$^5$ Helmholtz Research Academy Hesse for FAIR, Campus Riedberg, Max-von-Laue-Stra{\ss}e 12, D-60438 Frankfurt am Main, Germany

}
\ead{pflaumer@itp.uni-frankfurt.de}

\begin{indented}

\vspace*{16pt}
{\normalsize 
\item Asia-Pacific Symposium for Lattice Field Theory - APLAT 2020, 4\:-\,7 August 2020}
\vspace*{5pt}
\end{indented}

%----------------------------------------------------------------------------
\begin{abstract}%
We search for possibly existent bound states in the heavy-light tetraquark channels with quark content $ \bar{b}\bar{b}ud $, $ \bar{b}\bar{b}us $ and $ \bar{b}\bar{c}ud $ using lattice QCD. We carry out calculations on several gauge link ensembles with $ N_f=2+1 $ flavours of domain-wall fermions and consider a basis of local and non-local interpolators. Besides extracting the energy spectrum from the correlation matrices, we also perform a Lüscher analysis to extrapolate our results to infinite volume.
\end{abstract}
%----------------------------------------------------------------------------

\vspace{-0pc}
\section{\label{intro}Introduction}

One of the major challenges in QCD is to understand exotic hadrons. This is mainly motivated by experimental observations of mesons whose quantum numbers, masses or decays cannot be explained by ordinary quark-anti-quark pairs. A prominent example are the charged $ Z_b^\pm $ states with masses and decay channels suggesting the existence of a $ b\bar{b} $ pair, whereas the non-vanishing electrical charge indicates the presence of another light quark-antiquark pair \cite{Belle:2011aa}. 
These systems are theoretically extremely challenging to investigate due to several existing decay channels. Doubly-heavy tetraquark systems, which are simpler to study have quark content $ \bar{Q}\bar{Q}' qq' $, where $ Q,Q' \in \{b,c\} $ are heavy quarks and $ q,q' \in \{u,d,s\} $ represent light quarks.
Previous studies showed that this system forms a stable bound state in the heavy quark limit $ m_Q \rightarrow \infty $  \cite{Carlson:1987hh,Manohar:1992nd,Eichten:2017ffp}. Moreover, many investigations were carried out within quark models, effective field theories and QCD sum rules for physical $ b $ quark mass $ m_Q = m_b $ predicting a hadronically stable state \cite{Eichten:2017ffp,Karliner:2017qjm,Wang:2017uld,Park:2018wjk,Wang:2018atz,Liu:2019stu,Braaten:2020nwp}. Recently, Born-Oppenheimer investigations of a four-quark system containing a heavy antidiquark $ \bar{b}\bar{b} $ and a light diquark $ ud $ based on lattice QCD four-quark potentials predict a hadronically stable tetraquark in the $ I(J^P)=0(1^+) $ channel \cite{Bicudo:2012qt,Brown:2012tm,Bicudo:2015kna,Bicudo:2015vta,Bicudo:2016ooe} while in the $ I(J^P)=0(1^-) $ channel a resonance has been predicted.
\cite{Bicudo:2017szl}. More rigorous full lattice QCD studies confirmed the $ \bar{b}\bar{b}ud $ bound state and considered further heavy-light four-quark systems
\cite{Francis:2016hui,Francis:2017bjr,Francis:2018jyb,Junnarkar:2018twb,Leskovec:2019ioa,Hudspith:2020tdf, Mohanta:2020eed}. Here we report on our findings for the $ \bar{b}\bar{b}ud $, $ I(J^P)=0(1^+) $ channel \cite{Leskovec:2019ioa} and our progress concerning tetraquark systems with quark content $ \bar{b}\bar{b}us $ and $ \bar{b}\bar{c}ud $.

%----------------------------------------------------------------------------

\section{\label{sec:LatticeSetup}Lattice Setup}

All computations were performed using gauge link configurations generated by the RBC and UKQCD collaboration \cite{Aoki:2010dy, Blum:2014tka} with $ 2+1 $ flavours of domain-wall fermions \cite{Furman:1994ky,Kaplan:1992bt,Shamir:1993zy,Brower:2012vk} and the Iwasaki gauge action \cite{Iwasaki:1984cj}. We considered five ensembles (see Tab.\,\ref{tab:configurations} and Table~I in \cite{Leskovec:2019ioa}) differing in the lattice spacing $ a $ ($\approx 0.083\,\textrm{fm} \ldots 0.114\,\textrm{fm}$), lattice size (spatial extent $\approx 2.65\,\textrm{fm} \ldots 5.48\,\textrm{fm} $) and pion mass ($ \approx 139\,\textrm{MeV} \ldots 431\,\textrm{MeV}$). In this talk, we focus mainly on results obtained on the C005 ensemble highlighted in Tab.\,\ref{tab:configurations}. 
Smeared point-to all propagators were used for all quarks. The heavy b quarks were treated in the framework of Non-Relativistic QCD (NRQCD) \cite{Thacker:1990bm,Lepage:1992tx}, while the charm quark propagators were computed using a relativistic heavy quark action as described in \cite{Brown:2014ena}. In order to reduce the numerical cost of the computation, the all-mode-averaging technique was applied \cite{Blum:2012uh,Shintani:2014vja}.

\begin{table}[htb]
	\centering
	\begin{tabular}{cccccc} \hline \hline 
		Ensemble & $N_s^3 \times N_t$ & $a$ [fm] 	& $a m_{u;d}$ & $a m_{s}$ & $m_\pi$ [MeV]  \\ \hline
		C00078& $48^3 \times 96$ & $0.1141(3)$	& $0.00078$			  & $0.0362$           & $139(1)$ \\ \hline
		\rowcolor{linkBlue!10!white}C005 & $24^3 \times 64$	 & $0.1106(3)$	& $0.005\phantom{00}$ & $0.04\phantom{00}$ & $340(1)$ \\
		C01	 & $24^3 \times 64$	 & $0.1106(3)$  & $0.01\phantom{000}$ & $0.04\phantom{00}$ & $431(1)$ \\ \hline
		F004 & $32^3 \times 64$  & $0.0828(3)$  & $0.004\phantom{00}$ & $0.03\phantom{00}$ & $303(1)$ \\
		F006 & $32^3 \times 64$  & $0.0828(3)$	& $0.006\phantom{00}$ & $0.03\phantom{00}$ & $360(1)$ \\ \hline\hline
	\end{tabular}
	\caption{\label{tab:configurations}Gauge-link ensembles \cite{Aoki:2010dy, Blum:2014tka} used in this work. $N_s$, $N_t$: number of lattice sites in spatial and temporal directions; $a$: lattice spacing; $am_{u;d}$: bare up and down quark mass; $a m_{s}$: bare strange quark mass; $m_\pi$: pion mass. }
\end{table}

%----------------------------------------------------------------------------
\section{\label{sec:bbud}Hadronically stable Tetraquark $ \bar{b}\bar{b}ud $ in the $ I(J^P)=0(1^+) $ channel}

In a first step, we studied the energy spectrum of the doubly-bottom four quark system with two light quarks $ ud $ and quantum numbers $ I(J^P)=0(1^+) $ \cite{Leskovec:2019ioa}. The two lowest thresholds in this channel are the $ B B^\ast $ and the $ B^\ast B^\ast $ meson pairs, which differ only by $ 45\,\textrm{MeV} $.\\
In order to extract the energy spectrum, we considered two types of interpolating operators. On the one hand, we constructed local operators, where all four quarks are located at the same space-time position and the total momentum is projected to zero. On the other hand, we used so-called non-local operators, which describe two spatially separated mesons, each with definite momentum.\\
We included three local operators, namely two mesonic ones which create a $ BB^\ast $ ($ \op_1 $) and a $ B^\ast B^\ast $ ($ \op_2 $) structure, respectively, and a diquark-antidiquark operator ($ \op_3 $). Additionally, two non-local operators describing a $ BB^\ast $ ($ \op_4 $) and $ B^\ast B^\ast $ ($ \op_5 $) scattering state were added to the operator basis. The detailed construction of the operators is discussed in \cite{Leskovec:2019ioa}, but for completeness, we summarize them in Tab.\,\ref{tab:operators_bbud} using the following notation:
\begin{equation}
\begin{alignedat}{2}
&T_1(\Gamma_1, \Gamma_2) = \sum_{\xb} \, \bar{Q}_1 \Gamma_1 q_1(x) \, \bar{Q}_2 \Gamma_2 q_2(x), \, \quad 
T_2(\Gamma_1, \Gamma_2) = \sum_{\xb} \, \bar{Q}_1 \Gamma_1 q_2(x) \, \bar{Q}_2 \Gamma_2 q_1(x) \\
&D_1(\Gamma_1, \Gamma_2) = \sum_{\xb} \, \bar{Q}_1^a \Gamma_1 \bar{Q}_2^b(x) \, q_1^a \Gamma_2 q_2^b(x),\, \quad
D_2(\Gamma_1, \Gamma_2) = \sum_{\xb} \, \bar{Q}_1^a \Gamma_1 \bar{Q}_2^b(x) \, q_1^b \Gamma_2 q_2^a(x)\, \\
&M_1(\Gamma) = \sum_{\xb} \, \bar{Q}_1 \Gamma q_1(x), \quad M_2(\Gamma) = \sum_{\xb}\bar{Q}_2 \Gamma q_2(x), \quad
M_3(\Gamma) = \sum_{\xb} \, \bar{Q}_1 \Gamma q_2(x), \quad M_4(\Gamma) = \sum_{\xb}\bar{Q}_2 \Gamma q_1(x).
\end{alignedat}
\label{eq:Op_abbreviation}
\end{equation}
\begin{table}[h]
	\centering
	\begin{tabular}{|C{2.1cm}|C{2.25cm}|c|c|} \hline
		quark content \newline $ \bar{Q}_1\bar{Q}_2q_1q_2 $ & quantum numbers $ I(J^P) $ & local operators & non-local operators \\ \hline \hline
		\multirow{3}{*}{$ \bar{b}\bar{b}ud $}	& \multirow{3}{*}{$ 0(1^+) $}	
		& $ T_1(\gamma_j, \gamma_5) - T_2(\gamma_j, \gamma_5) $ 
		& $ M_1(\gamma_j) M_2(\gamma_5) - M_3(\gamma_j) M_4(\gamma_5) $ \\
		& & $\epsilon_{ijk} \left(T_1(\gamma_k, \gamma_j) - T_2(\gamma_k, \gamma_j)\right)$  & 
		$\epsilon_{ijk} \left( M_1(\gamma_k) M_2(\gamma_j) - M_3(\gamma_k) M_4(\gamma_j)\right) $ \\
		& & $ D_1(\gamma_j\C, \C\gamma_5) $ & \\ \hline

	\end{tabular}
	\caption{\label{tab:operators_bbud} List of operators that were considered for the $ \bar{b}\bar{b}ud $ correlation matrix. $ \C=\gamma_0 \gamma_2 $ denotes the charge conjugation matrix.}
\end{table}\\
We expect that the local operators will generate a state that predominantly overlaps with the ground state, i.e. describe the stable four-quark state of interest. Additionally the non-local operators are expected to have sizable overlap to the first excited state, i.e. a two meson state, which helps to isolate the ground state from the excitations.

\begin{figure}[h]
	\centering
	\begin{minipage}{0.58\textwidth}
		\includegraphics[width=\textwidth, trim = 15 15 15 15 ,clip]{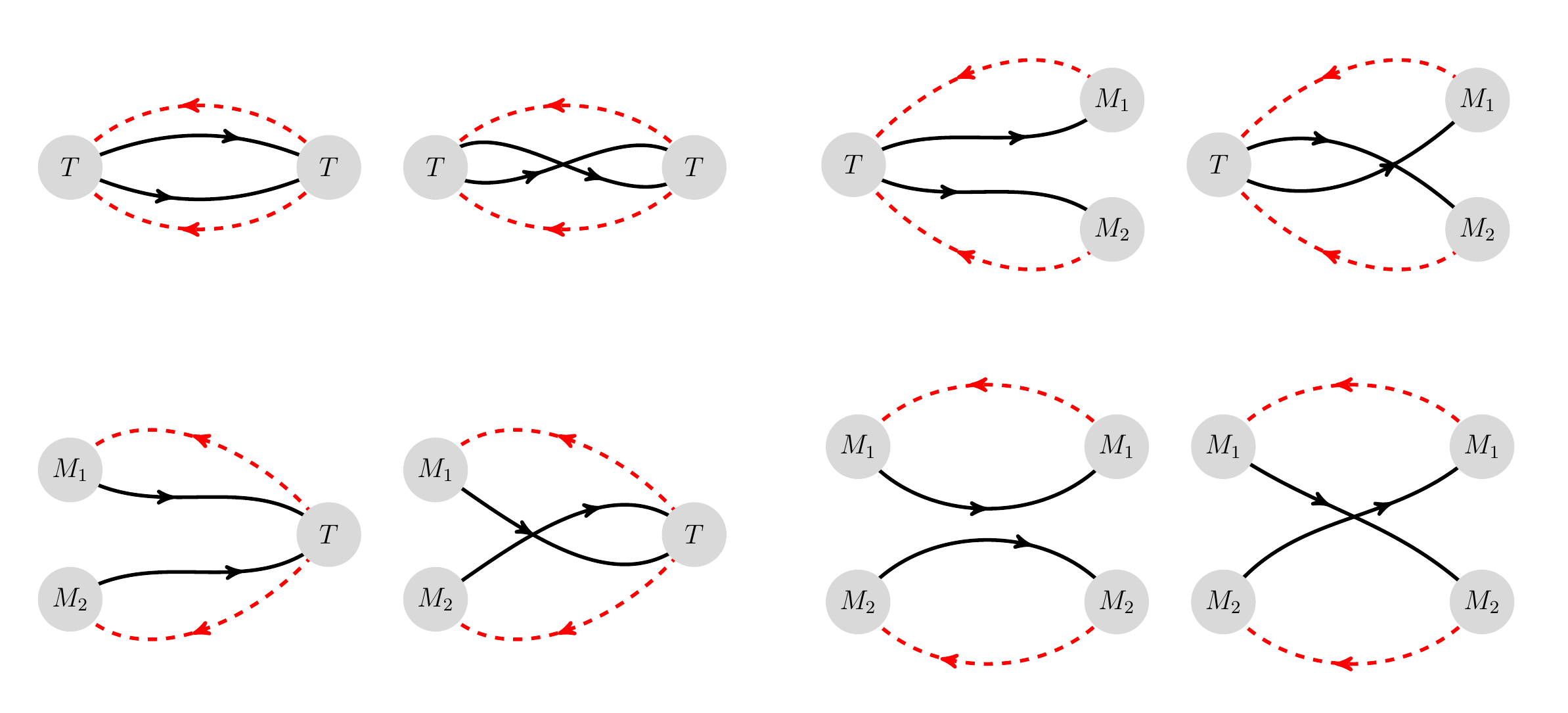}
		
	\end{minipage}
	\hspace{1pc}
	\begin{minipage}{0.38\textwidth}
		\caption{Schematic representation of Wick contractions for the elements of the correlation matrix. $ M_1 $ and $ M_2 $ represent
			the $ B $ and $ B^\ast $ mesons independently projected to zero momentum, while $ T $ represents local four-quark operators. The
			black lines represent $ b $ quark propagators and the red lines represent light quark propagators. \label{fig:WickContractions}}
	\end{minipage}
	
\end{figure}

\begin{figure}[h]
	\centering
	\begin{minipage}{0.58\textwidth}
		\includegraphics[width= \textwidth]{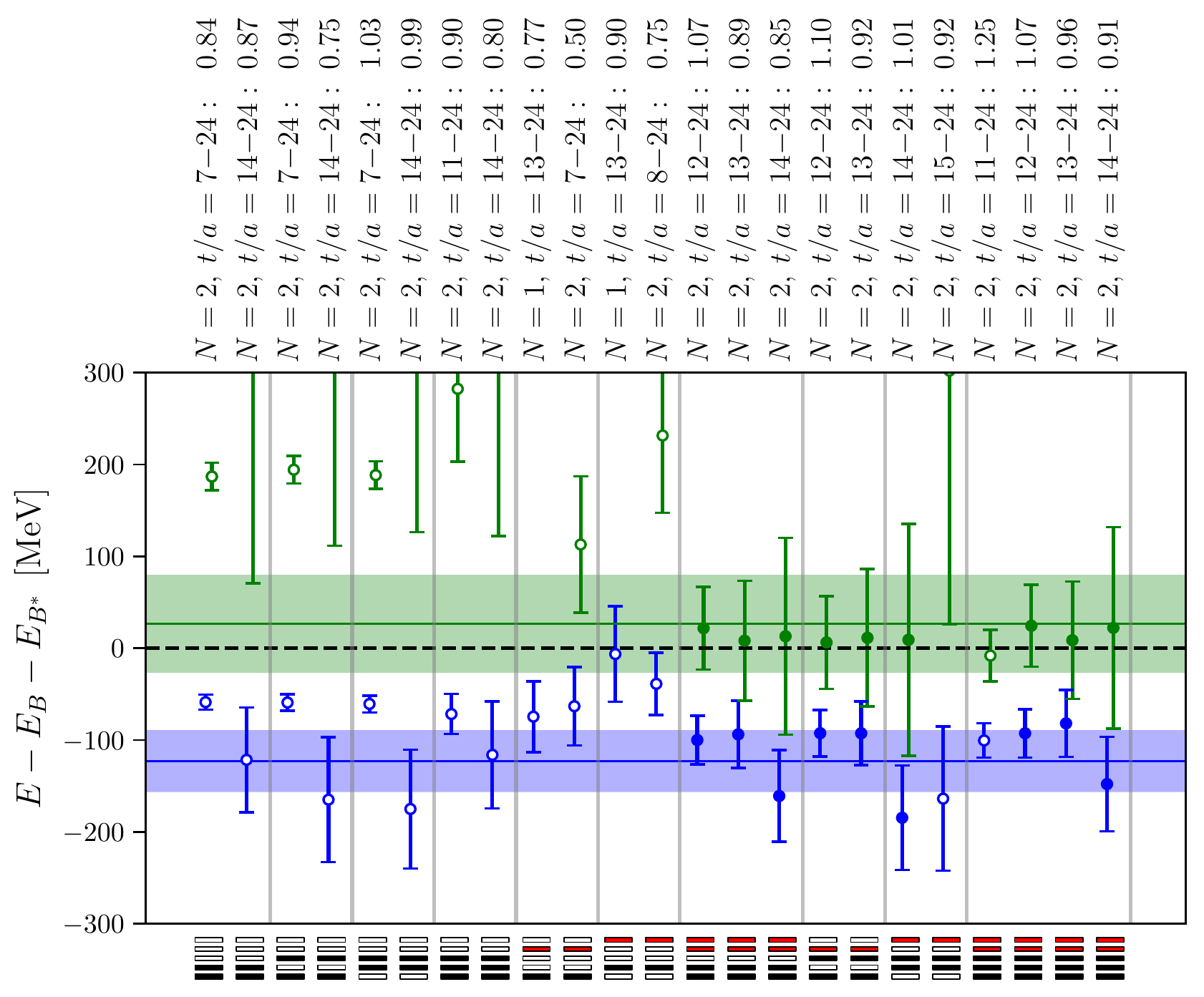}
	\end{minipage}
	\hspace{0.5pc}
	\begin{minipage}{0.385\textwidth}
		\caption{Results for the lowest two $ \bar{b}\bar{b}ud $ energy levels relative to the $ BB^\ast $ threshold, $ \Delta E_n = E_n - E_B - E_{B^\ast} $, as determined on ensemble C005 from several different fits. The five bars below each column indicate the interpolators used, a filled black box indicates a local operator included, a filled red box a scattering operator included. Above each column, we give the number of exponentials, the fit range, and the value of $ \chi^2/\textrm{d.o.f.} $ The shaded horizontal bands correspond to our final estimates of $ \Delta E_0 $ and $ \Delta E_1 $, obtained from a bootstrap average of the subset of fits that are shown with filled symbols. \label{fig:Fitresults_bbud}}
	\end{minipage}
\end{figure}

To determine the energy spectrum we considered the correlation matrix $ C_{j k}(t) = \langle \op_j(t) \op_k^\dagger(0)\rangle $ where $ \op_j $ and $\op_k $ represent one of the previously introduced interpolating operators. The corresponding schematic representation of the Wick contractions can be found in Fig.\,\ref{fig:WickContractions}. 
As we were using point-to-all propagators for the light quarks, we were restricted to correlation matrix elements with local operators at the source, which means that the resulting correlation matrix is a $ 5 \times 3 $ matrix. We applied a multi-exponential matrix fit to extract the energy levels, fitting
\begin{equation}
\label{eq:multiexp} C_{j k}(t) \approx \sum_{n=0}^{N-1}  Z_j^n Z_k^n \textrm{e}^{- {E_n} t}
\end{equation}
to the correlation matrix elements where $ E_n $ is the $ n $-th energy eigenvalue and $ Z_j^n = \langle n |\op_j^\dagger | \Omega \rangle $  are the overlaps of the trial states $  \op_j | n \rangle $ and the vacuum $ |\Omega\rangle $. We present the results for the two lowest energy levels for a large number of different fits in Fig.\,\ref{fig:Fitresults_bbud}. One observes that the energy levels become only stable if the non-local operators are included in the operator basis. Furthermore the energy levels are significantly lower compared to cases where non-local operators are absent. This shows the importance of scattering operators for the present study of the $ \bar{b}\bar{b}ud $ system. The ground state energy level is significantly below the $ BB^\ast $ threshold, while the first excited level is close to that threshold. This is a first indication that indeed a hadronically stable tetraquark state exists.\\

\begin{figure}[h]
	\centering
	\begin{minipage}{0.7\textwidth}
			\includegraphics[width= \textwidth]{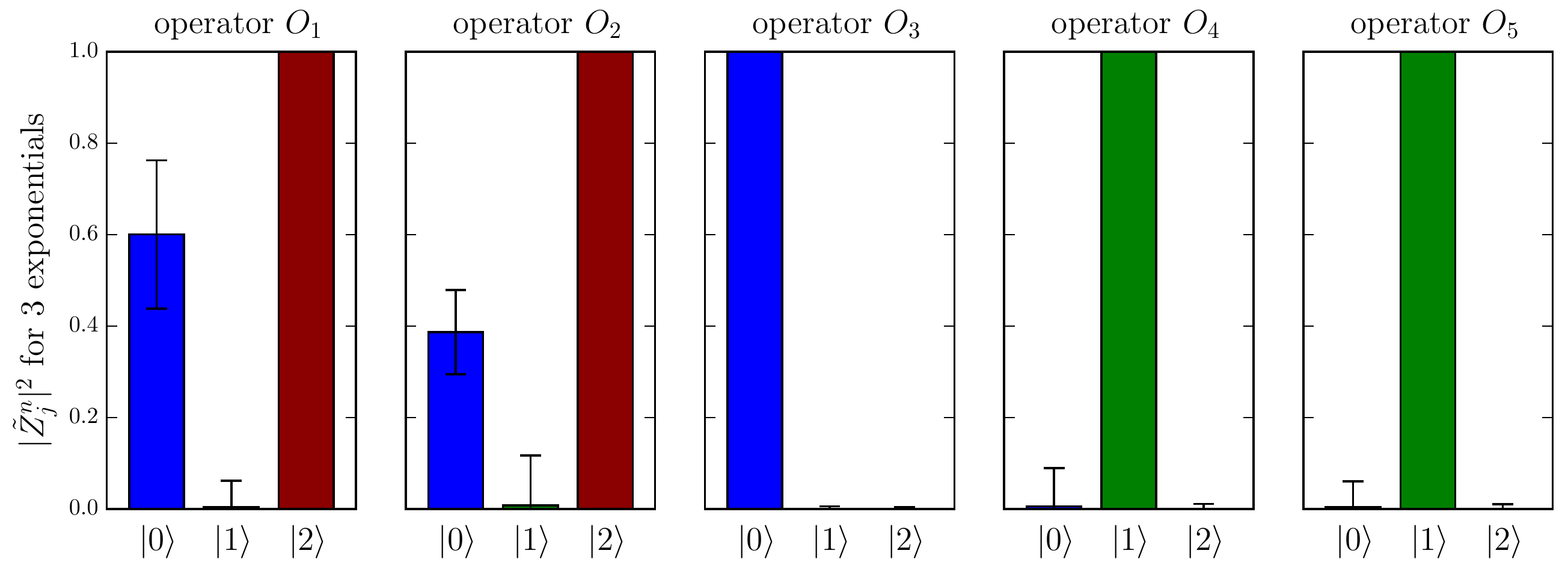}
	\end{minipage}
	\begin{minipage}{0.285\textwidth}
		\caption{The normalized overlap factors $|\tilde{Z}_j^n|^2$ as determined on ensemble C005 indicating the relative contributions of the energy eigenstates $ |n\rangle $ to the trial state $ \op_j^\dagger | \Omega\rangle $. \label{fig:overlap_factors}}
	\end{minipage}
\end{figure}

Additionally, we can gain certain information about the composition of the energy eigenstates $ |n\rangle $ by considering the overlap factors $ Z_j^n $. For a given operator index $ j $, the overlap factor $ Z_j^n $ introduced in Eq. \eqref{eq:multiexp} indicates the relative importance of an energy eigenstate $| n \rangle$, when the trial state $ \op_j^\dagger |\Omega\rangle $ is expanded in terms of energy eigenstates,
\begin{equation}
\op_j^\dagger | \Omega \rangle = \sum_{n=0}^\infty | n \rangle \langle n | \mathcal{O}_j^\dagger | \Omega \rangle = \sum_{n=0}^\infty  {Z_j^n} | n \rangle.
\end{equation}
If the overlap factor $ Z_j^m $ for a specific $ |m\rangle $ is dominant compared to all other $ Z_j^n $ with $ n\neq m $, this might indicate that the state $ \op_j^\dagger | \Omega \rangle $ is quite similar to $ |m \rangle $. In Fig.\,\ref{fig:overlap_factors} we present the overlap factors obtained by a 3-exponential fit for all five operators. The overlap factors $|\tilde{Z}_j^n|^2= |Z_j^n|^2 / \textrm{max}_m(|Z_j^m|^2)$ are normalized such that $ \textrm{max}_m(|Z_j^m|^2) = 1 $.
From Fig.\,\ref{fig:overlap_factors}, one can see that the trial states created by the non-local meson-meson scattering operators $ \op_4 $ and $ \op_5 $ have a large overlap to the first excited state $ |1\rangle $, which supports our assumption that the first excitation is a two-meson scattering state. Additionally, one can see that the trial state generated by the diquark-antidiquark operator $ \op_3 $ has significant overlap to the ground state $ |0\rangle $. This confirms our interpretation that the ground state represents the hadronically stable tetraquark. 

\begin{figure}[h]
	\centering
	\begin{minipage}{0.58\textwidth}
		\centering
		\includegraphics[width=0.75\textwidth]{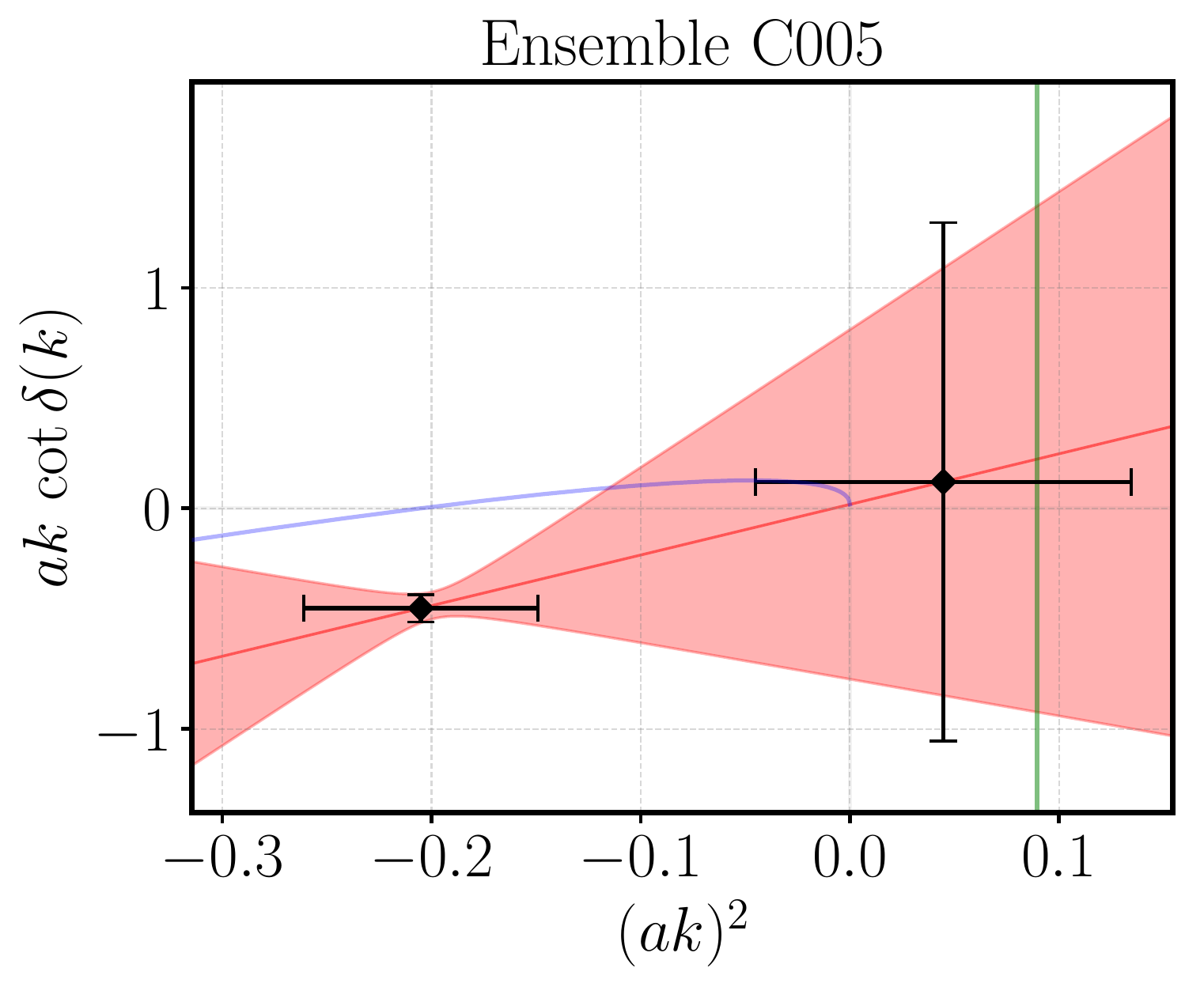}
	\end{minipage}
	\begin{minipage}{0.38\textwidth}
		\caption{Plot of the effective range expansion for ensemble C005. The red line corresponds to the ERE parametrization of $ ak \cot(\delta(k)) $. Below the $B B^*$ threshold (which is located at $k=0$), we also show the curves $ak \cot \delta(k) + |ak|$ (again using the ERE parameterization for $ak \cot \delta(k)$), whose lowest zero gives the binding momentum. The vertical green line indicates the inelastic $ B^\ast B^\ast $ threshold.\label{fig:ERE}}
	\end{minipage}
\end{figure}

The previously computed finite volume energy levels $ E_n $ can be related to the infinite volume scattering amplitude by applying Lüscher's method \cite{Luscher:1990ux}. Here, we used the two lowest energy levels shown in Fig.\,\ref{fig:Fitresults_bbud} and applied Lüscher's method to determine the $ BB^\ast $ S wave scattering amplitude. First, the finite volume scattering momenta $ k_n $ defined by 
\begin{equation}
E_n = E_B + \sqrt{m_{B,\,{\rm kin}}^2 + k_n^2} - m_{B,\,{\rm kin}} + E_{B^*} + \sqrt{m_{B^*\!,\,{\rm kin}}^2 + k_n^2} - m_{B^*\!,\,{\rm kin}},
\quad \textrm{with:  }\, m_{\rm kin} = \frac{\mathbf{p}^2-[E(\mathbf{p})-E(0)]^2}{2 [E(\mathbf{p})-E(0)]}
\end{equation}
are related to the infinite-volume phase shifts $ \delta_0(k_n) $ via
\begin{equation}
cot(\delta_0(k_n)) = \frac{2Z_{00}(1; (k_n L/2\pi)^2 )}{\pi^{1/2}k_n L}
\end{equation}
where $ Z_{00} $ is the generalized zeta function \cite{Luscher:1990ux}. 
The scattering amplitude is given by 
\begin{equation}
\label{eq:analyticcont} T_0(k) = \frac{1}{\cot\delta_0(k) -i}
\end{equation}
and can be parametrized by the effective range expansion (ERE)
\begin{equation}
k \cot \delta_0(k) = \frac{1}{a_0} + \frac{1}{2} r_0 k^2 + \mathcal{O}(k^4),
\end{equation}
where the two parameters $ a_0 $ and $ r_0 $ were determined using the finite volume energy levels $ \Delta E_0 $ and $ \Delta E_1 $ shown in Fig.\,\ref{fig:Fitresults_bbud}. We illustrate the ERE for ensemble C005 in Fig.\,\ref{fig:ERE}. 

\begin{figure}[h]
	\centering
	\begin{minipage}{0.58\textwidth}
		\centering
		\includegraphics[width=0.9\textwidth]{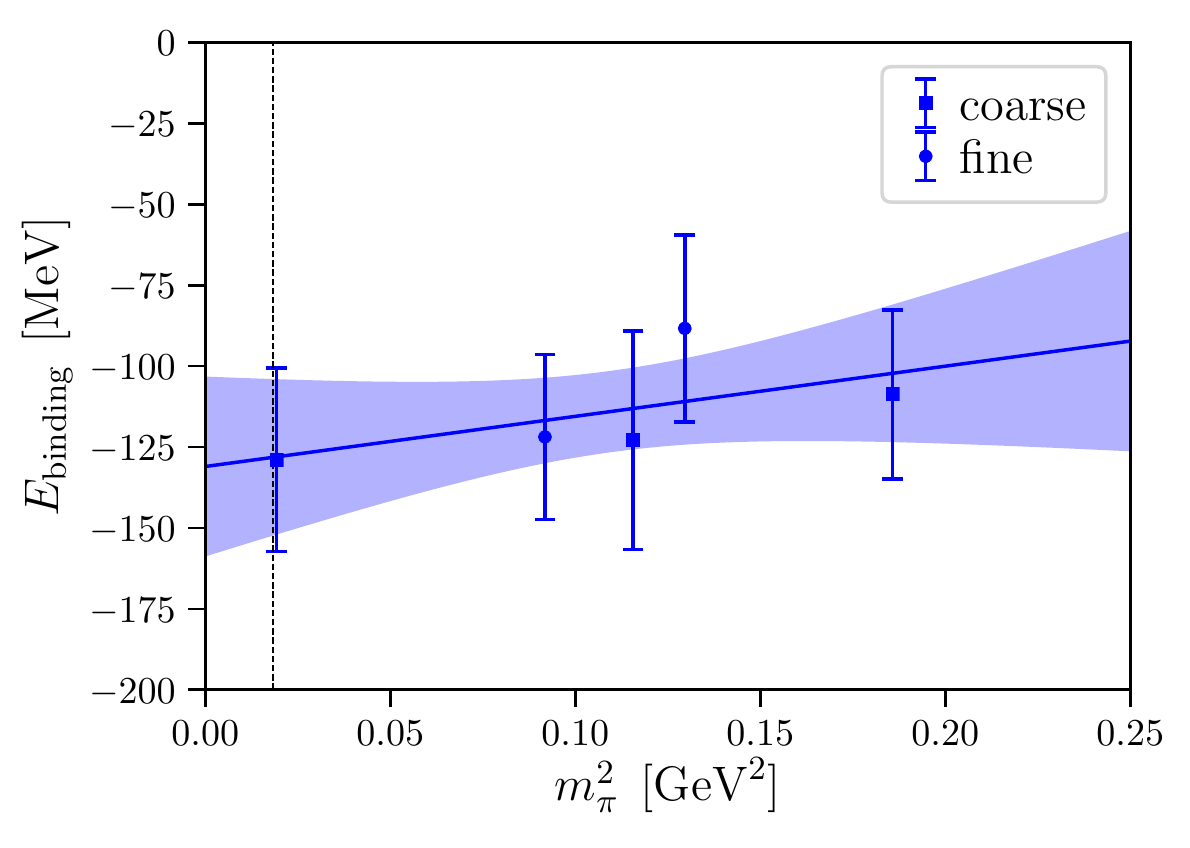}
	\end{minipage}
	\hspace{0.5pc}
	\begin{minipage}{0.38\textwidth}
		\caption{Extrapolation to the physical pion mass, indicated by the vertical dashed line. \label{fig:chiral_ex}}
	\end{minipage}
\end{figure}

Bound states appear as poles of the scattering amplitude below threshold. Combining this pole condition with the ERE yields 
\begin{equation}
-|k_{\rm BS}| = \frac{1}{a_0} - \frac{1}{2} r_0 |k_{\rm BS}|^2 ,
\label{eq:scatteringBS}
\end{equation}
where $ k_{\rm BS} $ is the bound state scattering momentum. Solving Eq.~\eqref{eq:scatteringBS} for $ |k_{\rm BS}| $, the binding energy is obtained via the NRQCD energy-momentum relation (see Sec.\,VI. in Ref.  \cite{Leskovec:2019ioa}). We found that the infinite volume mass of the bound state determined via the pole of the scattering amplitude is essentially identical to the finite volume ground state energy, which confirms the existence of a hadronically stable tetraquark.\\
Note, that all computations were repeated for all five ensembles discussed in Sec.\,\ref{sec:LatticeSetup}, which allowed us to perform an extrapolation to the physical pion mass. In Fig.\,\ref{fig:chiral_ex}, we present the final results for all ensembles as well as the extrapolation, for which we assume a quadratic pion mass dependence, $ E_\textrm{binding}(m_\pi) = E_\textrm{binding}(m_{\pi,\textrm{phys}}) + c(m^2_\pi - m^2_{\pi,\textrm{phys}}) $. Our final results for the tetraquark binding energy and mass are
\begin{equation}
E_\textrm{binding}(m_{\pi,\textrm{phys}}) = (-128 \pm 24 \pm 10)\,\textrm{MeV},\qquad m_\textrm{tetraquark}(m_{\pi,\textrm{phys}}) = (10476 \pm 24 \pm 10)\,\textrm{MeV}.
\end{equation}

%----------------------------------------------------------------------------
\section{Search for Bound States in the $ \bar{b}\bar{b}us $ and $ \bar{b}\bar{c}ud $ sector}

As discussed in the previous section, we were able to confirm the bound state $ \bar{b}\bar{b}ud $ in the $ 0(1^+) $ channel, which according to the literature is certainly the most promising candidate for a hadronically stable tetraquark. We are currently investigating further candidates, which have either a heavier light quark, i.e. an $ s $ quark instead of a $ d $ quark, or a lighter heavy quark, i.e. a $ \bar{c} $ quark instead of a $ \bar{b} $ quark.\\
In the case of $ \bar{b}\bar{b}us $, the most promising channel has again angular momentum $ J=1 $ and parity $ P=+ $, so we consider similar quantum numbers as for the $ \bar{b}\bar{b}ud $ system, namely $ I(J^P)=1/2(1^+)$. Previous lattice investigations of this state predict the ground state significantly below $ B_sB^\ast $ threshold \cite{Francis:2016hui,Junnarkar:2018twb}.\\
In the $ \bar{b}\bar{c}ud $ sector the picture is less clear. First, there are two channels in which a stable tetraquark could be expected, the $ 0(0^+) $ and the $ 0(1^+) $ channel. The first study found some indication for a $ \bar{b}\bar{c}ud $ bound state \cite{Francis:2017bjr,Francis:2018jyb}. This was, however, not confirmed by a more recent work \cite{Hudspith:2020tdf}.\\

\begin{figure}[h]
	\begin{minipage}{0.5 \textwidth}
		\includegraphics[width= \textwidth, trim = 5 5 5 146 ,clip]{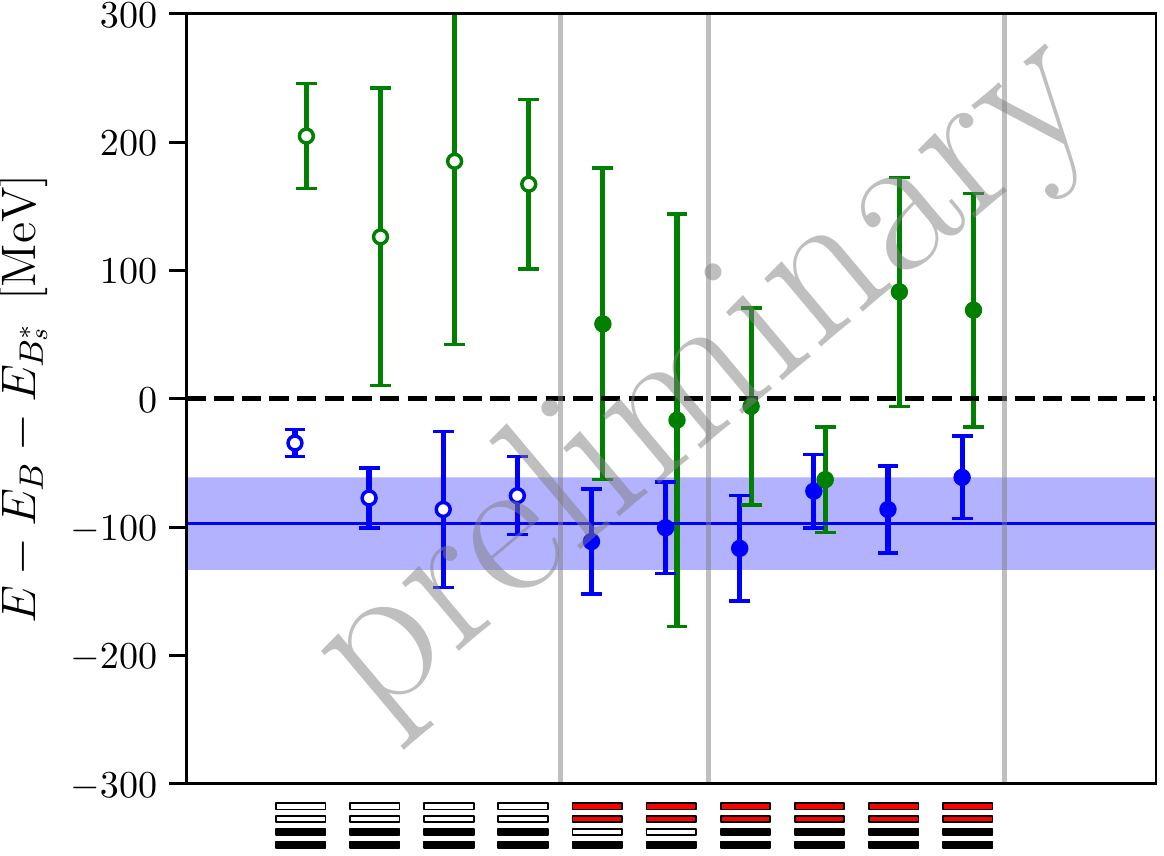}
	\end{minipage}
	\hspace{0.5pc}
	\begin{minipage}{0.5 \textwidth}
		\caption{Preliminary results for the lowest energy levels in the $ \bar{b}\bar{b}us $, $ 1/2(1^+) $ (top left), $ \bar{b}\bar{c}ud $, $ 0(0^+) $ (bottom left) and $ \bar{b}\bar{c}ud $, $ 0(1^+) $ (bottom right) sectors relative to the relevant threshold determined on ensemble C005 from several different fits. The bars below each column have the same meaning as discussed in Fig.\,\ref{fig:Fitresults_bbud}. \label{fig:Fitresults_bbus}}
	\end{minipage}
	\vskip 1pc
	\begin{minipage}{0.5 \textwidth}
			\includegraphics[width= \textwidth, trim = 5 5 5 146  ,clip]{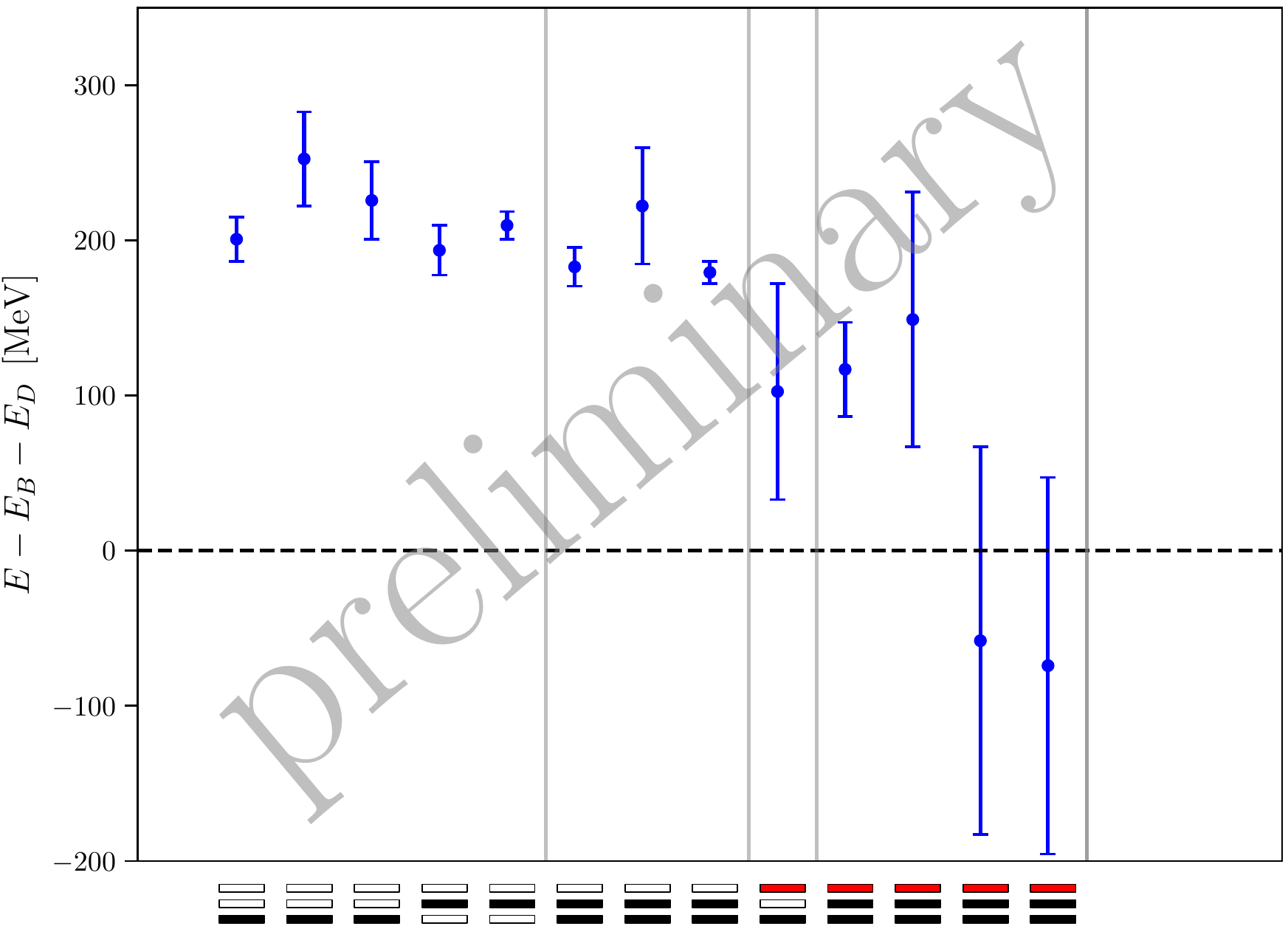}	
	\end{minipage}
	\begin{minipage}{0.5 \textwidth}
		\includegraphics[width= \textwidth, trim = 5 5 5 142, clip]{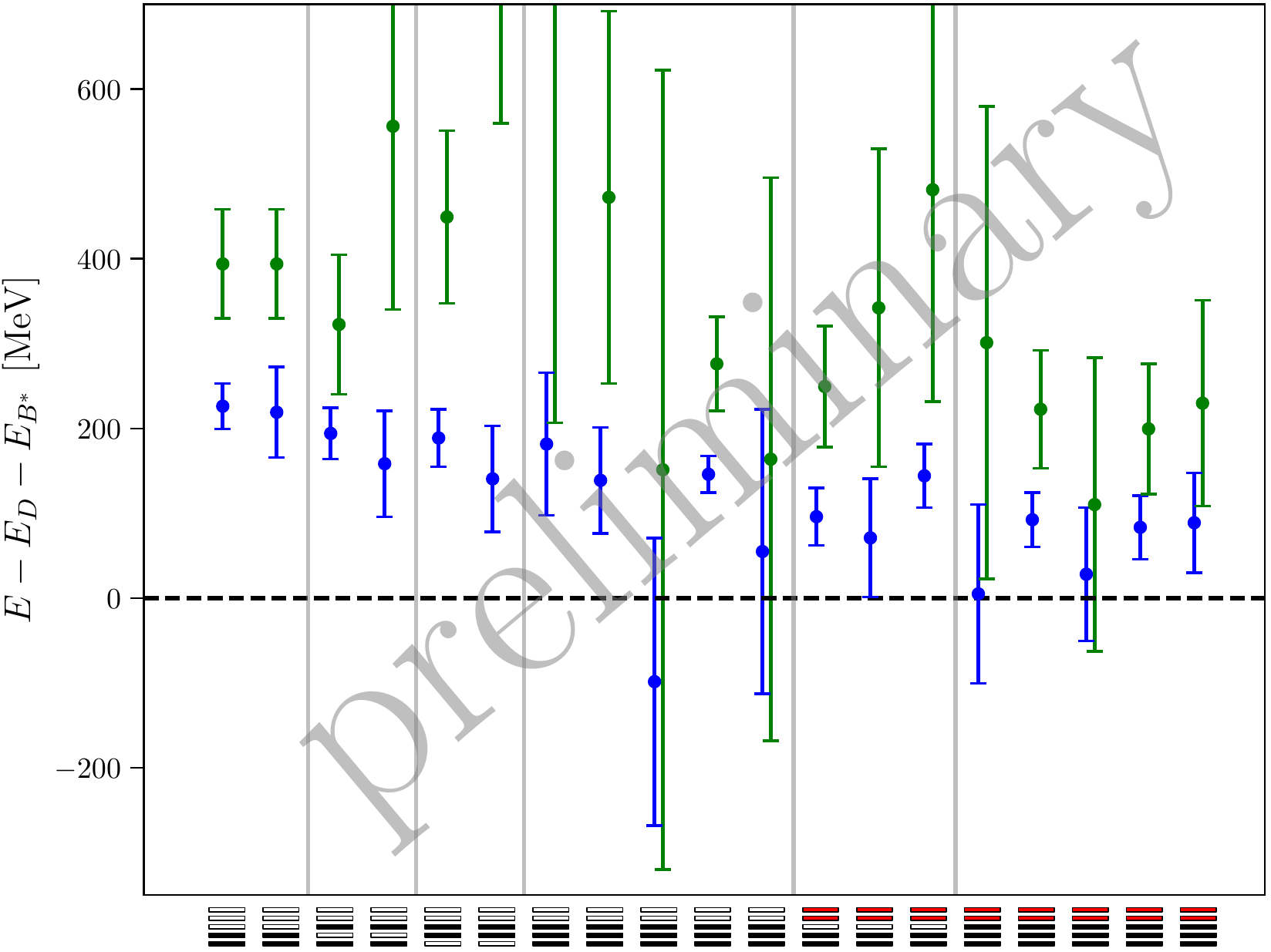}	
	\end{minipage}
\end{figure}

In our study of those tetraquarks, we utilize the same techniques as discussed for $ \bar{b}\bar{b}ud $ in Sec.\,\ref{sec:bbud}. In particular, we consider in all three cases local mesonic and diquark-antidiquark operators as well as meson-meson scattering operators. The employed interpolating operators are listed in Tab.\,\ref{tab:operators}, where we use again the notation given in Eq.~\eqref{eq:Op_abbreviation}. We expect that these operators are sufficient to extract the ground states correctly.\\
\begin{table}[h]
	\centering
	\begin{tabular}{|C{2.1cm}|C{2.25cm}|c|c|} \hline
		quark content \newline $ \bar{Q}_1\bar{Q}_2q_1q_2 $ & quantum numbers $ I(J^P) $ & local operators & non-local operators \\ \hline \hline
		\multirow{3}{*}{$ \bar{b}\bar{b}us $}	& \multirow{3}{*}{$ 1/2(1^+) $}	
		& $ T_1(\gamma_5, \gamma_j) - T_2(\gamma_5, \gamma_j) $ 
		& $ M_1(\gamma_5) M_2(\gamma_j) - M_3(\gamma_5) M_4(\gamma_j) $ \\
		& & $ \epsilon_{ijk} \left( T_1(\gamma_j, \gamma_k) - T_2(\gamma_j, \gamma_k) \right) $  & 
		$ \epsilon_{ijk} \left( M_1(\gamma_j) M_2(\gamma_k) - M_3(\gamma_j) M_4(\gamma_k) \right)$ \\
		& & $ D_1(\gamma_j\C,\C\gamma_5 ) $ & \\ \hline
		\multirow{3}{*}{$ \bar{b}\bar{c}ud $}	& \multirow{3}{*}{$ 0(1^+) $} 
		& $ T_1(\gamma_5, \gamma_j) - T_2(\gamma_5, \gamma_j) $ 
		& $ M_1(\gamma_5) M_2(\gamma_j) - M_3(\gamma_5) M_4(\gamma_j) $ \\
		& & $ T_1(\gamma_j, \gamma_5) - T_2(\gamma_j, \gamma_5)$  
		& $ M_1(\gamma_j) M_2(\gamma_5) - M_3(\gamma_j) M_4(\gamma_5) $ \\
		& & $ D_1( \gamma_j\C, \C\gamma_5) - D_2( \gamma_j\C, \C\gamma_5)$ & \\ \hline
		\multirow{2}{*}{$ \bar{b}\bar{c}ud $}	& \multirow{2}{*}{$ 0(0^+) $} 
		& $ T_1(\gamma_5, \gamma_5) - T_2(\gamma_5, \gamma_5) $
		& $ M_1(\gamma_5) M_2(\gamma_5) - M_3(\gamma_5) M_4(\gamma_5) $ \\
		& & $ D_1(\C\gamma_5, \C\gamma_5)  - D_2(\C\gamma_5, \C\gamma_5) $ & \\ \hline
	\end{tabular}
\caption{\label{tab:operators} List of operators considered for the corresponding correlation matrix. Note that for $ \bar{b}\bar{b}us $ we assume an approximate $ \textrm{SU}(3) $ flavor symmetry as strange and light quarks are almost massless compared to the $ b $ quark.}
\end{table}\\
For $ \bar{b}\bar{b}us $ we found that certain elements of the corresponding correlation matrix are strongly correlated if all local operators are included in the operator basis. Thus, we restricted the operator basis by generating a new set of local interpolating operators via linear combination of the original ones. The contribution of the operator $ \op_j $ to the new operator $ \op_n' $ is determined by the eigenvector component $ v_j^n $, where $ \vec{v^n} $ is the eigenvector obtained by solving a $ 3\times 3 $ generalized eigenvalue problem (GEP). We use the eigenvectors corresponding to the two lowest energy levels to construct two new operators. Thus, the trial state $ \op_1'|n\rangle $ should overlap most strongly with the ground state, while $ \op_2'| n \rangle $ will be more similar to the first excited state.
The new operators read consequently
\begin{equation}
	\op'_n = \sum_{j=1}^{3}\, v_j^n\, \op_j,
\end{equation}
where the eigenvector components can be found in Tab.\,\ref{tab:ev_components}. The scattering operators were not changed, i.e. we used a $ 4\times 2 $ matrix in case of $ \bar{b}\bar{b}us $.
\begin{table}[h]
	\centering
	\begin{tabular}{|c|c|c|c|}\hline
	$ v_j^n $	& $j=1 $	& $ j=2 $ 	& $ j=3 $ \\ \hline
	$ n=1 $		& $ +0.72 $	& $ +0.57 $ & $ +0.39 $ \\ \hline		
	$ n=2 $		& $ +0.66 $	& $ -0.75 $ & $ +0.05 $ \\ \hline
	\end{tabular}
\caption{\label{tab:ev_components} Eigenvector components calculated via a GEP for the $ 3x3 $ correlation matrix including all local operators.}
\end{table}
For $ \bar{b}\bar{c}ud $, $ 0(0^+) $ and $ \bar{b}\bar{c}ud $, $ 0(1^+) $, we used a $ 3\times 2 $ and a $ 5\times 3 $ correlation matrix, respectively. 
The energy levels were extracted for all cases in the same way as described in Sec.\,\ref{sec:bbud} and we present the results in Fig.\,\ref{fig:Fitresults_bbus}. Note that for $ \bar{b}\bar{c}ud $, $ 0(0^+) $ we only show results for the lowest energy level, as the first excitation could not be properly isolated from higher excitations.\\

We found in the $ \bar{b}\bar{b}us $ channel a ground state energy level somewhat below the $ B^\ast B_s $ threshold, which indicates a bound state. Moreover, the first excited state seems to be close to the threshold and, thus, might be consistent with a meson-meson scattering state. Averaging different fits on the C005 ensemble, we generated a crude estimate of the binding energy $ E_{\bar{b}\bar{b}us, \textrm{binding}}( m_\pi = 340\, \textrm{MeV} ) \approx (-100\pm 40) \,\textrm{MeV} $, which is in reasonable agreement with results from Ref. \cite{Francis:2016hui,Junnarkar:2018twb}.\\

Assuming that the extracted energies for $ \bar{b}\bar{c} ud $ shown in Fig.\,\ref{fig:Fitresults_bbus}, in particular those, where non-local meson-meson scattering operators were used, reflect the ground state energies, there is no indication for a bound state, neither for $ 0(0^+) $ nor for $ 0(1^+) $. In both cases, the ground state seems to be close to the threshold, which suggests that the lowest energy level is rather a scattering state than a bound four-quark state. This interpretation is further supported by observing a significant decrease of the lowest energy level when non-local meson-meson scattering operators were considered in the fits. This indicates a rather large overlap of the ground state to the scattering trial states. In particular for $ 0(0^+) $ this is expected from lattice QCD studies based on static-static-light-light potentials, where the $ 0(0^+) $ potential is significantly less attractive than the $ 0(1^+) $ potential (see Ref. \cite{Bicudo:2016ooe} , Eqs. (12b) and (16b)).

\FloatBarrier
\vspace{-1pc}
\section*{Acknowledgments}

We thank Antje Peters for collaboration in the early stages of this project. We thank the RBC and UKQCD collaborations for providing the gauge field ensembles.
L.L.\ acknowledges support from the U.S. Department of Energy, Office of Science, through contracts DE-SC0019229 and DE-AC05-06OR23177 (JLAB).
S.M.\ is supported by the U.S. Department of Energy, Office of Science, Office of High Energy Physics under Award Number D{E-S}{C0}009913.
M.W.\ acknowledges support by the Heisenberg Programme of the Deutsche Forschungsgemeinschaft (DFG, German Research Foundation) - project number 399217702. Calculations  on the  GOETHE-HLR  and  on  the  FUCHS-CSC  high-performance  computers  of the Frankfurt University were conducted for this research.  We would like to thank HPC-Hessen,funded by the State Ministry of Higher Education, Research and the Arts, for programming advice.
This research used resources of the National Energy Research Scientific Computing Center (NERSC), a U.S.\ Department of Energy Office of Science User Facility operated under Contract No.\ DE-AC02-05CH11231. This work also used resources at the Texas Advanced Computing Center that are part of the Extreme Science and Engineering Discovery Environment (XSEDE), which is supported by National Science Foundation grant number ACI-1548562.

% ********************

%%%%%%%%%%%%%%%%%%%%%%%%%%%%%%%%%%%%%%%%%%%%%%%%%%%%%%%%%%%%%%%%%%%%%%%%%%%%%
\end{document}